\documentclass[aps,prb,floatfix,twocolumn,showpacs]{revtex4}
\usepackage{latexsym}
\usepackage{graphicx}
\usepackage{dcolumn}
\newcommand{\figwidth}{2.725 in}
\begin{document}
\title{Quantum nematic as ground state of a 
two-dimensional electron gas in a magnetic field}
\author{Quoc M. Doan$^{1}$ and Efstratios Manousakis$^{1,2}$}
\affiliation{$^{1}$Department of Physics and MARTECH, 
Florida State University, 
Tallahassee, FL 32306-4350, USA and \\
$^{2}$Department of  Physics, University of Athens,
Penipistimiopolis, Zografos, 157 84 Athens, Greece.}  
\date{\today}
\begin{abstract}
We study the ground state of a nematic phase of the two-dimensional
electron gas at filling
fraction $\nu = 1/2$ using a variational wavefunction
having Jastrow pair-correlations of the form $\Pi_{i < j}(z_i-z_j)^2$
and an elliptical Fermi sea. 
Using the Fermi hypernetted chain approximation we find
that below a critical value
of the broken symmetry parameter, the nematic phase is energetically
favorable as compared to the isotropic state 
for the second excited Landau level. We also find that 
below a critical value of the layer ``thickness'' parameter $\lambda$
(and in the actual materials) the quantum nematic is energetically
favorable relative to the stripe ordered Wigner crystal phase. 
\end{abstract}
\maketitle
During the past two decades, the quantum Hall effect (QHE) 
has been one of the most
intriguing research topics in condensed matter\cite{chakraborty}. 
More recently, the measurements of Lilly {\it et. al.}\cite{lilly} 
and Du {\it et. al.}\cite{du} reveal a strong
anisotropic behavior of transport properties of electrons for the 
half-filled Landau level (LL)
system under strong magnetic field and at very low temperature.  The anisotropy
commences at the second excited LL and persists up to the sixth excited
LL. The sudden 
exhibition
of large anisotropies of resistivities in clean two-dimensional electron gas
(2DEG) suggests that there is an unknown underlying microscopic origin for
this kind of spontaneously symmetry breaking.

These experimental findings prompted several
interesting theoretical proposals which attempt to explain the
observed anisotropic behavior of the half-filled LL system.
First of all, these anisotropic transport properties are consistent
with already predicted stripe and bubble charge-density-wave phases 
found\cite{koulakov,moessner} by means of Hartree-Fock calculations 
of the 2DEG. 
However, Fradkin and Kivelson\cite{fk} suggested 
that the anisotropic transport might be due to a stripe nematic phase of the
2DEG in a high magnetic field.
This point of view was investigated further by  Fradkin 
{\it et al.}\cite{fradkin} 
where a model for the nematic phase in a symmetry breaking field 
was studied using Monte Carlo
simulation. The results of the Monte Carlo simulation provide
a good fit of the experimental data
of Lilly {\it et al}.\cite{lilly}. 
This simulation suggests that the nematic phase might
be a good candidate to explain the anisotropic behavior observed in 
Ref.~\onlinecite{lilly} and Ref.~\onlinecite{du}. 
Furthermore, by deriving a long-wavelength
elastic theory of the quantum Hall smectic state,
Wexler and Dorsey\cite{dorsey} have estimated the 
transition temperature
from an isotropic to nematic phase to be of the order of
$ 200 mK$.  Later, Cooper {\it et al.}\cite{cooper}
by applying an in-plane magnetic field in 2DEG samples which show
the above anisotropy in transport, give further support for the
possible presence of such a quantum nematic phase.

In the composite fermion theory given by Jain\cite{jain}, the 
fractional quantum Hall effect (FQHE) is interpreted
as the integer quantum Hall effect  of composite Fermions. 
Furthermore, Halperin {\it et al.}\cite{halperin} developed a
theory of half-filled LL system, which is the case of our interest, 
as a compressible Fermi
liquid. Rezayi and Read\cite{rezayi} proposed a ground state wave
function for the half-filled LL system having the Jastrow-Slater form as
follows:
\begin{eqnarray}
\Psi \left( {\vec {r}_1 ,\vec {r}_2 ,...,\vec {r}_N } \right)=\hat
{P} \prod\limits_{j<k}^N {(z_j -} z_k )^2 e^{-1/4 \sum_{k=1}^N
\left| {z_k } \right|^2 } \nonumber \\
\times \det \left| {\varphi _{\vec {k}} (\vec
{r}_i )} \right|,
\label{rezayi}
\end{eqnarray}
where  $\varphi_{\vec {k}} (\vec {r}_i)$ are two-dimensional (2D) 
plane-wave states, and
$\hat {P}$ is the LL projector operator. 
Here $z_j = x_j+ i y_j$ is the complex 2D coordinate of the $j$ electron.
This wavefunction is a Jastrow correlated  Slater
determinant with Jastrow part similar to the Laughlin state.

Ciftja and Wexler\cite{wexler} used the 
Fermi-hypernetted-chain (FHNC) approximation 
to study a broken
 rotational state of the half-filled LL where the symmetry breaking parameter
 was introduced in the correlation part of the wavefunction and the
single particle determinant of the wavefunction was characterized by
a circular Fermi sea.  

The ground state wavefunction for the 
nematic state proposed in Ref.~\onlinecite{oganesyan} has the same
form as the wavefunction given by Eq.~\ref{rezayi}, however, the
single particle momenta form
an elliptical Fermi sea  as opposed to the circular Fermi sea. 
In this paper we will use this wavefunction 
 to study the nematic phase. The broken symmetry parameter in our problem 
is the ratio $\alpha=k_1/k_2$ of the major $k_{1}$ and minor $k_{2}$ axis of 
the elliptic Fermi sea. 
We will study the nematic state of the half-filled LL
system using the variational approach and we will employ the Fermi hypernetted 
chain approximation\cite{ciftja,fr,krotscheck,manousakis}. 
We adopt the ansatz for the ground state 
of the nematic state proposed by Oganesyan {\it et al.}\cite{oganesyan} 
as trial wavefunction in our variational
calculations.  Namely, we investigate whether or not this state,
in which the anisotropy is due to an elliptical Fermi sea, 
can be energetically favorable relative to the isotropic state 
and the stripe ordered Wigner crystal at high LL.
We find that the nematic phase can be stabilized against the
isotropic case beyond the
second excited LL. In addition, we have compared the energy of the
nematic state to that obtained by a self-consistent Hartree-Fock
calculation\cite{cote1} of the stripe and bubble states of
the 2DEG. We find that there is a transition between
the stripe ordered ground state and the nematic phase
as a function of the parameter $\lambda$ of the Zhang-Das 
Sarma interaction\cite{dassarma}, i.e., the layer finite thickness.  
In particular,
for the case of the materials\cite{lilly,du} in question, we find that
the ground state corresponds to the nematic state.


The FHNC formalism for Fermi systems was introduced and developed in
Refs.~\onlinecite{fr,krotscheck,manousakis,ciftja}. A significant
advantage of FHNC over the variational Monte Carlo is that 
FHNC does not suffer from  finite-size effects. In the current problem,
where we need to estimate small energy
differences, the role of finite-size effects may be significant.
The main idea of this method is to expand the pair
 distribution function in powers of the density of the system
and it works well for a low density system.
The variational Monte Carlo becomes advantageous
 at relatively high density, as in the case of liquid
$^3He$, where the role of the elementary diagrams and other
non-trivial many-body correlations need to be included.\cite{manousakis}

In order to apply the FHNC formalism, first, we assume that the
unprojected wavefunction is good
approximation to the real wavefunction for the LLL. Moreover,
it is known\cite{trivedi}
that the projection operator almost eliminates the high LL
components of the wavefunction. The potential energy of the high LL 
can be expressed\cite{ciftja}
via the pair distribution function of the LLL using the single mode
approximation discussed in Ref.~\onlinecite{macdonald}, namely:
\begin{eqnarray}
\bar V_L = \frac{\rho}{2}\int[g(r)-1]V^{(L)}_{eff}(r)d^2r,\label{potential}
\end{eqnarray}
where the effective potential $V^{(L)}_{eff}(r)$ at Landau level L
is the convolution of the effective  Zhang-Das Sarma (ZDS) 
interaction\cite{dassarma},
$V(r) = {e^2}/\epsilon\sqrt{r^2+\lambda^2}$,
with the $L$-order Laguerre polynomials; namely, it is the Fourier transform of
${\tilde V}^{(L)}_{eff}(q) = ({2\pi e^2})/ {\epsilon q} \exp(-\lambda
q)[L_L(q^2/2)]^2$.
In the above formula $\lambda$ is a length scale
which characterizes the confinement of the electron wavefunction
in the direction perpendicular to the heterojunction.\cite{dassarma} 

Our calculation proceeds as follows:
First, we will calculate the pair distribution functions for isotropic
  and nematic states with different values of the anisotropic parameter 
for the LLL using the FHNC approximation. Second, 
the interaction energies will be calculated via the pair
  distribution functions by using the single-LL approximation, i.e., via
Eq.~\ref{potential}. Next, the kinetic energy is evaluated for the 
isotropic and different
  nematic states. The energy values of 
the isotropic state are compared with anisotropic
  states for the lowest, first and second excited LL to find out if
the nematic state becomes energetically favorable. 
Finally, we will carry out a self-consistent
Hartree-Fock calculation\cite{cote1,dorsey2} for the more general case 
where $\lambda$ can be non-zero and we will compare the energy of the
nematic, isotropic and stripe ordered  Wigner crystal. 

In the FHNC technique, each term in the expansion is
represented as graphical diagrams with well-defined topological
rules. There are nodal, composite and elementary diagrams.
In the FHNC/0 approximation, which neglects the elementary
diagrams, the pair distribution function is obtained by solving 
the FHNC integral equations given in Ref.~\onlinecite{manousakis} 
for the case of polarized 
liquid  $^3He$. These equations require as input, (a) the pair correlation 
(or Jastrow) factor, which in our case is given as
$f^2({r})=\exp(u({ r}))$ with $u({r})=4 ln(r)$, and (b) the statistical
exchange factor $l(r)$, which for a 2D Slater
determinant is given by: $l(r)=C(k_F r)$,
where $C(x)\equiv 2 J_1(x)/x$, $J_1(r)$ is the first order
Bessel function, and $k_F$ is the Fermi momentum of the isotropic
state. Alternatively for the anisotropic state having an elliptic
Fermi surface with major and minor axes $k_1$ and $k_2$, we find
$l({\bf r})=C(X)$, with $X=\sqrt{(k_1 x)^2+(k_2 y)^2}$
where $x$ and $y$ are the coordinates of ${\bf r}$.  

Since the pseudo-potential $u(r)$ has a long-range logarithmic form, we 
follow the standard procedure used in Ref.~\onlinecite{wexler} and 
Ref.~\onlinecite{chakraborty} to separate it into a 
short-range and a long-range part.
This leads to a new set of FHNC equations for the
short range nodal and composite functions which are solved 
iteratively using a combination 
of momentum space and real space approach.\cite{ciftja}


\begin{figure}[htp]
\vskip 0.3 in
\begin{center}
\includegraphics[width=\figwidth]{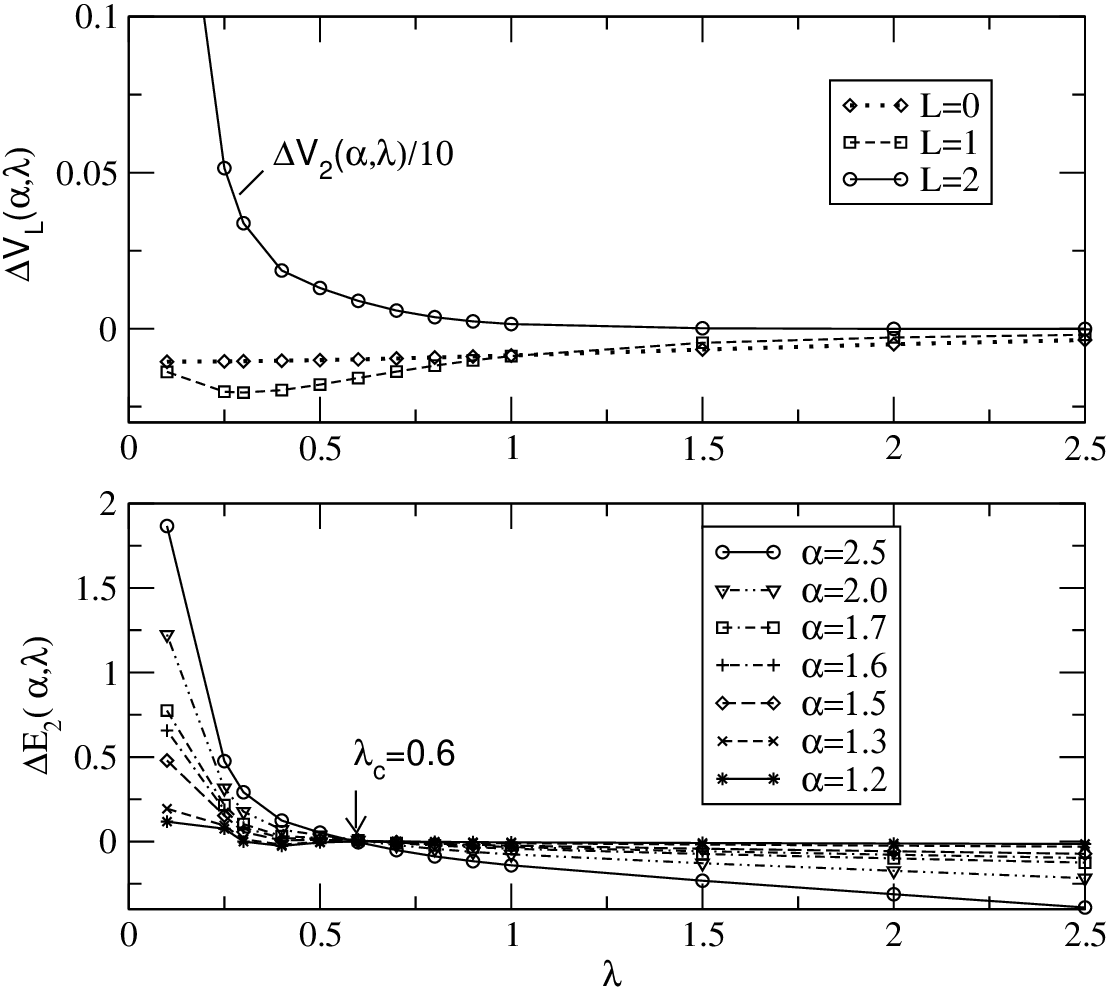}
\caption{\label{fig1}
Top: The potential energy difference $\Delta V_L(\alpha,\lambda)$ 
as a function of $\lambda$ for anisotropy parameter 
$\alpha=2.5$ for $L=0$, $L=1$, and $L=2$ (scaled down by a factor of 10).
 Bottom: The total energy difference $\Delta E_2(\alpha,\lambda)$ 
as a function of $\lambda$ for various values of the
Fermi sea anisotropy parameter $\alpha$ for the second excited LL.}
\end{center} 
\end{figure}

We are interested in the potential and total energy difference 
between the isotropic ($\alpha=1$) and the anisotropic case ($\alpha>1$),
namely,
$\Delta V_L(\alpha,\lambda) = V_L(1,\lambda)-V_L(\alpha,\lambda),$
and $\Delta E_L(\alpha,\lambda) = E_L(1,\lambda)-E_L(\alpha,\lambda)$,
as a function of $\lambda$  and the anisotropy parameter $\alpha$ and 
for $L=0$, $L=1$ and $L=2$.
In the interesting region, 
the energy difference between the nematic and the isotropic state  can be
small relative to the energy scale $e^2/l_0$,
so high accuracy may be required. 
The effective potential for high LL and for
small values of $\lambda$ changes rapidly at small distances and oscillates
at large distances. We have used an adaptive mesh to incorporate 
these multi-scale oscillations accurately for up to $L=2$. 
For higher Landau levels it becomes increasingly
more difficult to carry out an accurate calculation due to 
the fact that these oscillations become increasingly more rapid.

Our calculated pair distribution function for the isotropic
accurately reproduces the pair distribution function reported
in Ref.~\onlinecite{ciftja}.
The potential energy difference $\Delta V_L(\alpha,\lambda)$ 
for various values of $\alpha$ is calculated and is shown in 
Fig.~\ref{fig1}(top), in units
of $e^2/(\epsilon l_0)$ (where $l_0 = \sqrt{\hbar c /e B}$).
Notice that for the case of the  LLL 
and for the first excited LL, $\Delta V_L(\alpha,\lambda) < 0$, 
i.e., the isotropic state is energetically favorable for all values 
of $\lambda$ and $\alpha$.
However, as it is illustrated in Fig.~\ref{fig1}(top), for the case of the
second excited LL, $\Delta V_2(\alpha,\lambda)>0 $,
for all values of $\alpha$ and for some range 
of the parameter $\lambda$, the anisotropic state can be energetically 
favorable provided that the energy loss due to the anisotropy of the
Fermi surface is not larger than the potential energy gain.

\begin{figure}[htp]
\vskip 0.2 in
\begin{center}
\includegraphics[width=\figwidth]{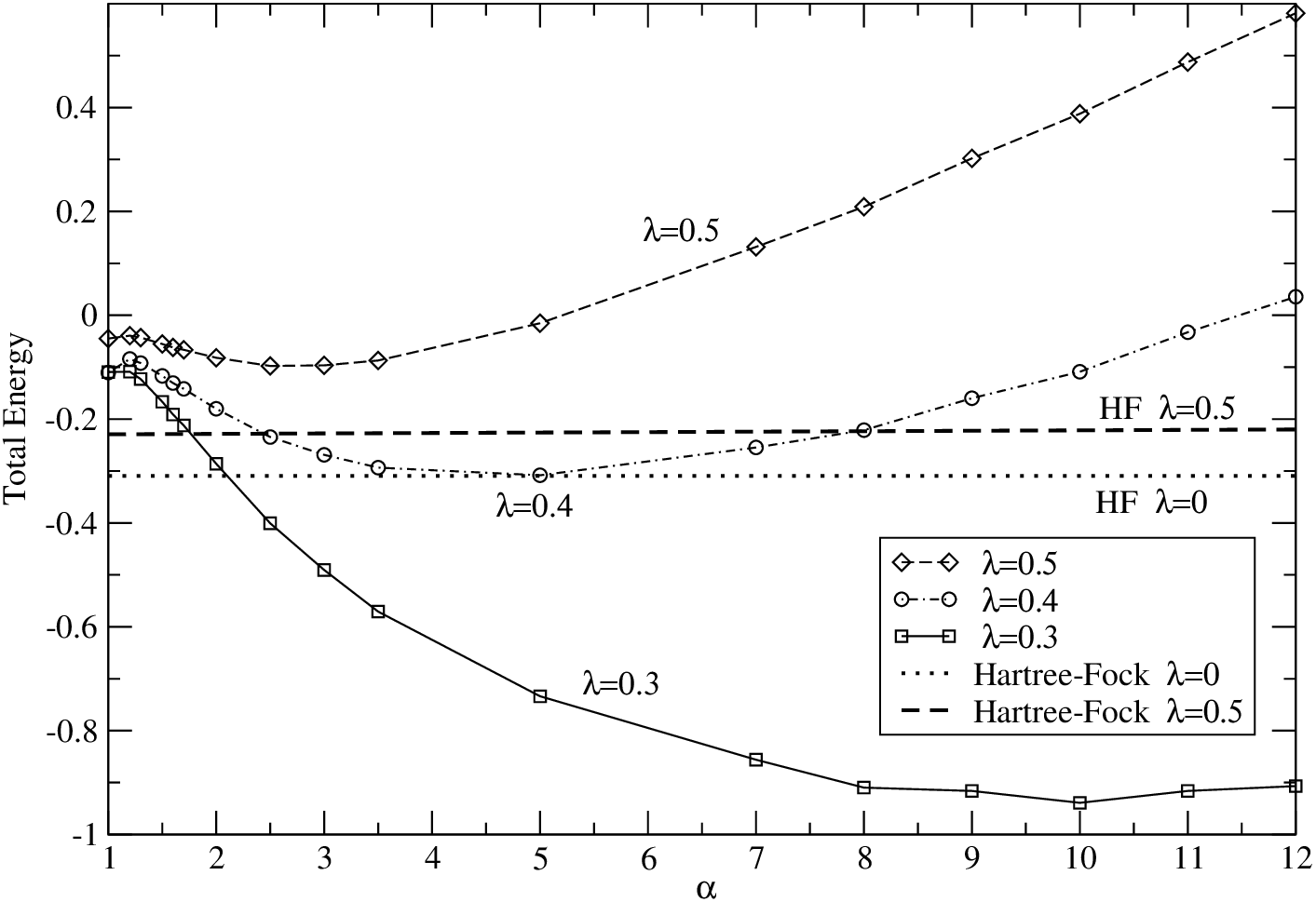}
\caption{\label{fig2}
The total energy as function of the anisotropy parameter 
$\alpha$ for various values of $\lambda$.}
\end{center} 
\end{figure}

\begin{figure}[htp]
\vskip 0.2 in
\begin{center}
\includegraphics[width=\figwidth]{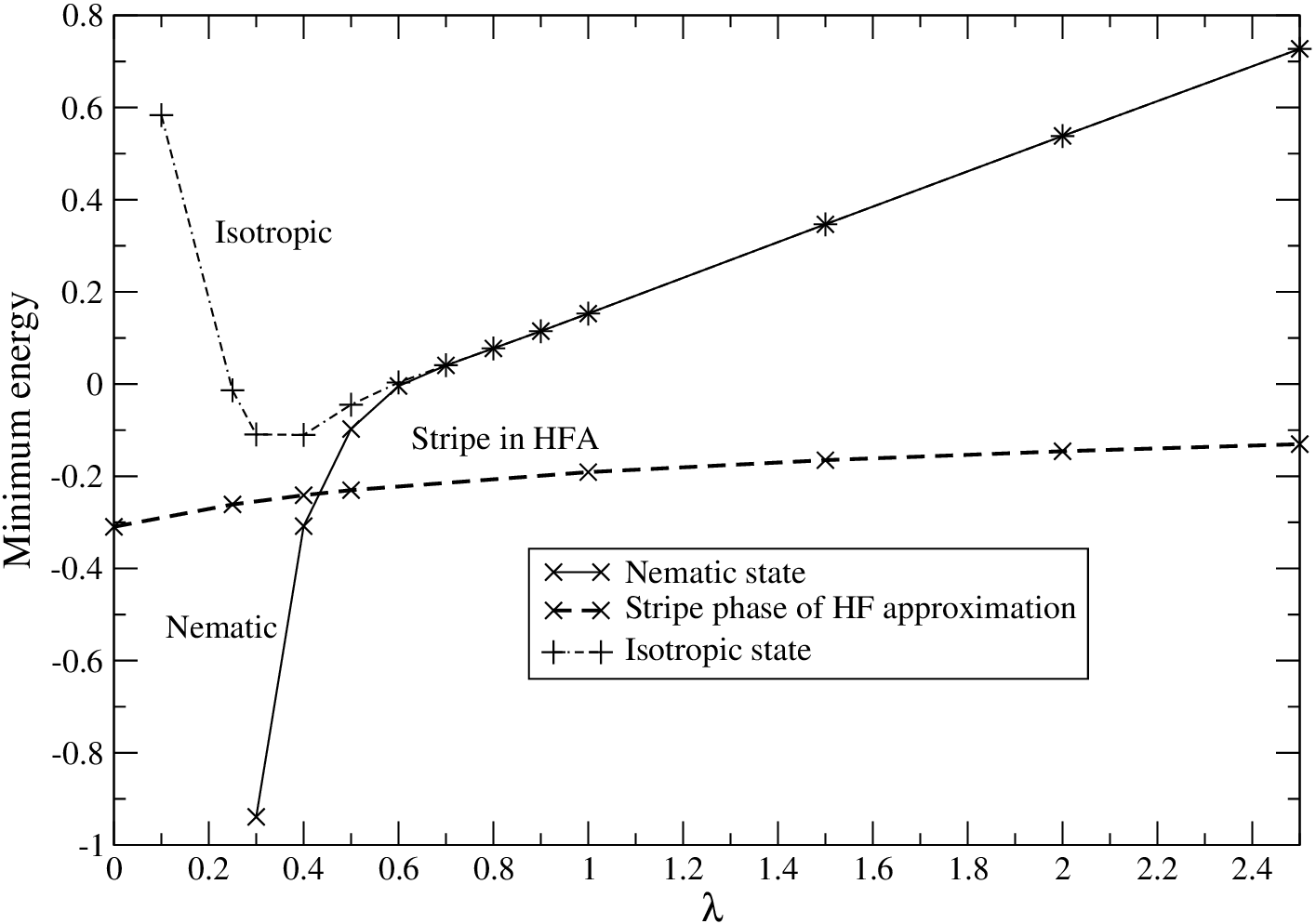}
\caption{\label{fig3}
The optimum values of the total energy of the nematic and isotropic 
state as a function 
$\lambda$ is compared with the results of the Hartree-Fock approximation
obtained for the same values of $\lambda$.}
\end{center} 
\end{figure}

In the single-LL approximation\cite{macdonald} the kinetic
energy of the isotropic state is quenched. We can estimate the kinetic
energy difference between the isotropic and the anisotropic 
case by ignoring the Landau level projection operator and, thus,
 writing the wave function as $\Psi = F \Phi$ where $\Phi$ is the
non-interacting Slater determinant and $F$ the Jastrow part.
The kinetic energy contains terms in which the operator 
$(\nabla - \vec A)^2$  acts on $F$. This term gives the same contribution
of $\hbar \omega_c/2$ in both isotropic and anisotropic case. 
Therefore, the main difference,
coming from the term $|F|^2 \Phi^* \nabla^2\Phi$,
is due to the difference in 
shape of the Fermi sea. This leads to the following kinetic energy
difference between isotropic and anisotropic Fermi sea:
$\Delta K \simeq - {{\hbar^2 k_F^2}\over {4m^*}}
{{(1-\alpha)^2} \over {2\alpha}}$.
In Fig.~\ref{fig1}(bottom) we present the total energy difference 
$\Delta E_2(\alpha,\lambda)$ between the anisotropic and the isotropic
state. Notice that the nematic state is energetically favorable relative to the
isotropic below $\lambda_c\simeq 0.6$.

Hartree-Fock energies of the stripe and bubble charge-density-wave energy
have already been reported;\cite{cote1,dorsey2} however, they
are only available for the case of $\lambda=0$.
In order to compare the energy of the nematic state with that of the
ordered stripe and the bubble states for a non-zero value of
$\lambda$, we carried
out a Hartree-Fock calculation using the method outlined in 
Refs.~\onlinecite{cote1,dorsey2}.
In Fig.~\ref{fig2} the total energy (apart from a common constant value
of $\hbar \omega_c/2$) is compared with the results of our Hartree-Fock 
calculation for finite values of $\lambda$.
In addition, in Fig.~\ref{fig3}, the minimum energy with respect to
the anisotropy parameter $\alpha$ is compared to the minimum
Hartree-Fock energy value with respect to the uniaxial anisotropy 
parameter $\epsilon$ (the lattice constants of the uniaxial
Wigner crystal are given in terms
of $\epsilon$ as  $a_1=\sqrt{3}a /2 \sqrt{1-\epsilon}$ and
$a_2=\sqrt{1-\epsilon} a/2$). 
Notice that there is a critical of $\lambda$, namely, $\lambda_c\simeq 0.4$
below which the nematic phase is energetically favored.

A value for the parameter $\lambda$ can be 
estimated using the calculation presented in Ref.~\onlinecite{dassarma}.
Using the value of the 2D electron density for these materials\cite{lilly}
we find that $\lambda \sim 62 \AA$. 
This corresponds to a value of $\lambda \sim 0.34$ in units 
of the magnetic length $l_0$ ($l_0=181 \AA$ for the value of $B \sim 2T$ 
which corresponds to the second excited LL in the experiments of 
Refs.~\onlinecite{lilly,du}).  
This value of $\lambda$ is  less than the 
critical value $\lambda_c\simeq 0.4$ below which the nematic phase
is energetically favorable as compared to the stripe ordered phase 
(see Fig.~\ref{fig3}).
Therefore, we conclude that our calculation suggests that 
for the case of the  2DEG in the heterojunctions used 
in Refs.~\onlinecite{lilly,du}  the quantum nematic  
state\cite{fk,fradkin,oganesyan} may be energetically lower than
the stripe ordered or bubble phases. Furthermore, it is interesting to 
probe this transition from the nematic to stripe to isotropic
by either experimentally altering the value of $\lambda$ or 
indirectly by means of an in-plane field.



\begin{figure}[htp]
\begin{center}
\vskip 0.2 in
\includegraphics[width=\figwidth]{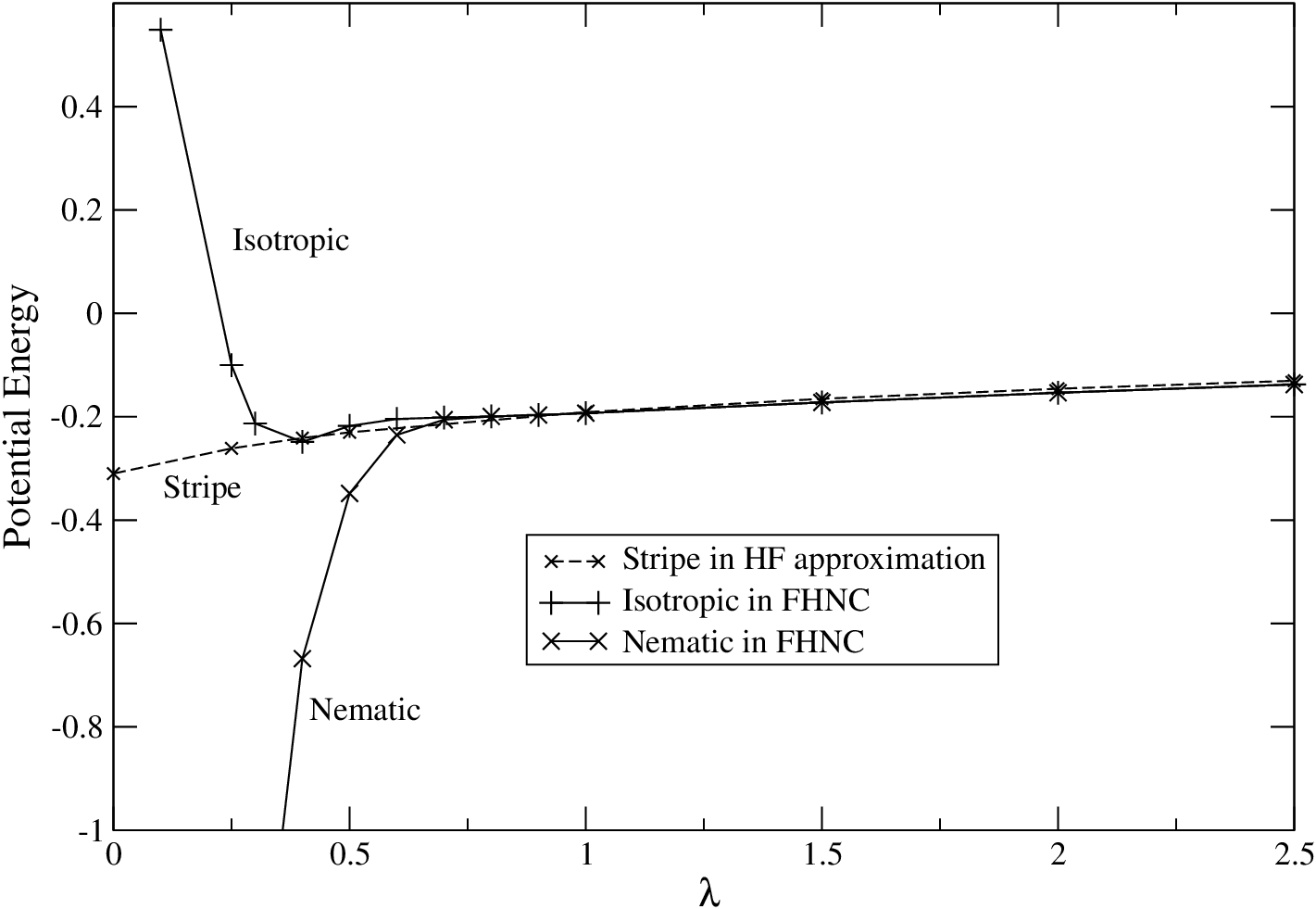}
\caption{\label{fig4}
The potential energy 
of the nematic and isotropic states as a function 
$\lambda$ is compared with the results of the Hartree-Fock approximation.}
\end{center} 
\end{figure}

So far, because  the Hartree Fock calculations predict a  
stripe state, the observed anisotropy in transport was taken as a signature 
of a stripe ordered state. This state breaks both translational invariance 
in one  direction and rotational invariance. The results  
of Haldane and Rezayi and Yang\cite{yang} are also usually interpreted 
as a stripe state.  
However, the systems which can be done with such an approach are very small, 
and, in addition, toroidal boundary conditions were used, which break 
rotational  invariance. Therefore, because of these limitations
it cannot be discerned if the true ground state is a stripe or a nematic. 
In our calculation we find that the
optimum nematic phase corresponds to an anisotropy of $\alpha \sim 10$
near the physically realized value of the parameter $\lambda$. 
This implies that a nematic state with such large anisotropy 
cannot be distinguished from
the stripe state in systems with only twelve electrons.\cite{yang}

There are two possible sources of systematic error in the present 
calculation.
First, the use of the FHNC/0 approximation to evaluate the distribution
function, where the contribution of the elementary diagrams is
neglected. This approximation works very well in low-density
systems, i.e., where the average interparticle distance
is large compared to a  hard core diameter. In the present problem
such a condition is not clearly fulfilled as there is only a soft
core of size $\lambda$. The second source of error is the fact that 
we have neglected the
projection operator and assumed that the unprojected wave function
given by Eq.~\ref{rezayi} is a good approximation to the
lowest LL. In order to address these concerns, in Fig.~\ref{fig4} we 
compare only the potential energy 
of the nematic and isotropic states  with the results of 
the Hartree-Fock approximation
obtained for the same values of $\lambda$. Notice that for values of
$\lambda > 0.6$ the results of the FHNC and the HF calculation are
almost identical. Moreover, the results of the FHNC calculation for the
isotropic state agree very well with those of the HF calculation
for all values of $\lambda > 0.3$. This is an indication
that the  energy difference between the nematic phase and that of the
isotropic state and the stripe state below $\lambda \simeq 0.5$ may
not be an artifact of the difference in the treatment of the two states
(i.e., the difference between HF and FHNC approximations) 
but rather due to the fact that
the nematic state for long-range interactions is energetically favorable
for at least the second excited LL.



We would like to thank L. W. Engel, E. Fradkin and S. A. Kivelson for useful
discussions. This work was supported by NASA under Grant No. NAG-2867.

\end{document}